\documentclass[twocolumn,secnumarabic,amssymb, nobibnotes, aps, pra,showpacs,preprintnumbers]{revtex4-1}
\usepackage{amssymb}
\usepackage{array}
\usepackage{graphicx}
\usepackage{amsmath}
\usepackage{amsthm}
\usepackage{amsopn}
\def\uno{\mbox{1 \kern-.59em {\rm l}}}

\def\beq{\begin{equation}}
\def\eeq{\end{equation}}
\def\bea{\begin{eqnarray}}
\def\eea{\end{eqnarray}}

\begin{document}

\title{Mimetic f(T) Teleparallel Gravity And Cosmology}%

\author{Behrouz Mirza}
\email{b.mirza@cc.iut.ac.ir}
\author{Fatemeh Oboudiat}
\email{f.oboudiat@ph.iut.ac.ir}
\affiliation{Department of Physics,
Isfahan University of Technology, Isfahan 84156-83111, Iran}

\begin{abstract}
We formulate mimetic theory in $f(T)$ teleparallel gravity where $T$ is torsion scalar. It is shown that the construction of the mimetic theory in the teleparallel gravity requires the vierbeins to be left unchanged and  the conformal transformation performed on the Minkowski metric of the tangent space. It is further argued that  the conformal degree of freedom in this teleparallel mimetic theory becomes dynamical and mimics the behavior of cold dark matter. We also show that it is possible to employ the Lagrange multipliers method to formulate the mimetic theory in an $f(T)$ theory without any auxiliary metric. The mimetic $f(T)$ theory is examined by the method of dynamical system and it is found that there are five fixed points representing inflation, radiation, matter, mimetic dark matter and dark energy dominated eras in the theory if some conditions are satisfied. We examined the power-law model and its phase trajectories with these conditions.
\end{abstract}


\maketitle

\newpage
\section{Introduction}
One of the most important issues in recent cosmology is the dark energy/matter problem \cite{dark matter}. An interesting model has been recently proposed  based on the conformal invariant extension of the theory which leads to new degree of freedom. The extra degree of freedom mimics the behavior of cold dark matter; hence, the designation mimetic theory \cite{Mimetic}. The role of the initial conditions for such a dark matter (DM) is discussed in \cite{initial-conditions}. In \cite{imperfect-DM1} and \cite{imperfect-DM2} the connection of the mimetic gravity and imperfect DM is represented. This theory  has been extended to explain inhomogeneous dark energy \cite{lagrange1}. While the cosmology  of the theory is investigated in \cite{lagrange4}, the unimodular mimetic cosmology, null energy condition violation, and $f(R)$ mimetic dark matter are addressed  in \cite{uni}, \cite{nec}, and \cite{frmimetic}, respectively. In mimetic $f(R)$ gravity, dark energy oscillations, unified inflation-dark energy evolution of the universe, aspects of late time evolution of the universe, Schwarzschild de sitter black holes and unimodular mimetic $f(R)$ inflation is discussed respectively in \cite{deofr}, \cite{inacfr}, \cite{lateevofr}, \cite{Schwarzschildfr} and \cite{unimodularfr}. Mimetic theory with higher derivatives has ghost degrees of freedom \cite{ghost}. However, these ghosts leads to a very slow instability \cite{stabilitymimetic}. The caustics along with the constraints from the solar system is studied in \cite{caustics} while connections to Einstein Aether were stressed in \cite{Einstein-Aether1} and \cite{Einstein-Aether2}. Lagrange multipliers method is studied in \cite{lagrange1,lagrange2,lagrange3,lagrange4} while in \cite{gauge-invariance} it is discussed how the gauge invariance of the mimetic gravity can be used to arrive to the formulation with Lagrange multipliers. Unified description of dark energy and dark matter in mimetic model is specified in \cite{unidedm}.
\\Teleparallel Gravity (TG) \cite{teleparallel} is an alternative formulation of General Relativity (GR) which,
 instead of the torsionless Levi-Civita connection, employs the curvatureless Weitzenb\"{o}ck connection \cite{weits} in the action.  While general relativity is assumed to be a geometric theory, teleparallel gravity is considered to be a gauge one \cite{gauge}. Surprisingly, however, the equations of motion of both  theories are equivalent so that  the theory is sometimes  called the Teleparallel equivalent of General Relativity (TEGR). Similar to the generalization of GR to $f(R)$ gravity, a straightforward generalization of TG is the $f(T)$ theory, where $T$ is replaced with $f(T)$ in the action \cite{ft1,ft2}. Compared to $f(R)$, $f(T)$ has the advantage of having  second order field equations in lieu of the fourth order ones in the $f(R)$ theory but it is not possible to fix some of the vierbeins by gauge symmetry as there is no local Lorenz invariance. On the other hand, $f(T)$ theory can explain inflation \cite{inflation} and capture acceleration of the universe \cite{ft1}. There are some problems with propagation and time evolution in $f(T)$ gravity theories \cite{propagation}. In \cite{observ}, the model parameters are constrained by observational data while  the advantage of adding a scalar field to the theory is observed in \cite{scalar}. The likelihood of wormholes and phantom divide crossing are considered in \cite{wormhole} and \cite{phantomcross}, respectively. Finally, the Noether symmetry is studied in \cite{Neother} and the number of degrees of freedom in $f(T)$ theory is studied in \cite{df}.
\\The method of dynamical systems is a useful tool for studying the whole dynamics of a theory near critical points called fixed points \cite{rev fix}. This method is employed to study $ f(T)$ theory for special forms of $f(T)$ function in \cite{fix ft} and for general case in \cite{Oboudiat}.
\\In this paper, we propose a new extension of the mimetic generalization of the TG and $f(T)$ theories. For this purpose, we first explain the basis of the mimetic formulation in General Relativity in Sections II and that of TG in Section III. The mimetic $f(T)$ Teleparallel Gravity is then investigated using the variational method in Section IV and the Lagrange multipliers in Sections V. Dynamical system analysis is used to study the mimetic $f(T)$ theory in Section VI. Finally, Section VII presents a summary and the conclusions.
\section{Mimetic model}
In the mimetic model, the metric is written in terms of an auxiliary metric and a scalar field appearing through its first derivative \cite{Mimetic}. As a consequence, the conformal degree of freedom of the action is isolated in a covariant way. It is interesting that the scalar field reveals  itself in the equations of motion. The physical metric may be parameterized as follows:
\bea
g_{\mu\nu}=\widetilde{g}_{\mu\nu}(\widetilde{g}^{\alpha\beta}\partial_\alpha\phi\partial_\beta\phi)\label{mimetic}
\eea
where, $\widetilde{g}_{\mu\nu}$ is the conformal extension of the metric $g_{\mu\nu}$ and $\phi$ is a scalar field. It is obvious from (\ref{mimetic}) that the scalar field satisfies the constraint below:
\bea
g^{\mu\nu}\partial_\mu\phi\partial_\nu\phi =1\label{constraint}
\eea
The equations of motion are derived by varying  the action with respect to the metric. The action is of the following form in General Relativity (GR) :
\bea
I=\int d^4x\sqrt{-g(\widetilde{g}_{\mu\nu},\phi)}\left[-\frac{1}{2}R(g_{\mu\nu}(\widetilde{g}_{\mu\nu},\phi))+\mathcal{L}_m\right]\label{graction}
\eea
where, $g$ is the trace of the metric, $R$ is the Ricci scalar, and $\mathcal{L}_m$ is the Lagrangian density of matter taken here to be  $8\pi G=1$. Instead of varying the action with respect to $g_{\mu\nu}$, it is varied  in the GR version of the mimetic model with respect to $\widetilde{g}_{\mu\nu}$ and $\phi$. This yields the following equations of motion:
\bea
&&(G^{\mu\nu}+T^{\mu\nu})-(G+T)g^{\mu\alpha}g^{\nu\beta}\partial_{\alpha}\phi\partial_{\beta}\phi=0\label{motion1}\\
&&\nabla_{\mu}((G+T)\partial^{\mu}\phi)=0\label{motion2}
\eea
where, $G^{\mu\nu}$ and $T^{\mu\nu}$ are Einstein and stress energy tensors and $G$ and $T$ are their traces. $\nabla_{\mu}$ denotes the derivative with respect to the physical metric $g_{\mu\nu}$. Eqs. (\ref{constraint}), (\ref{motion1}), and (\ref{motion2}) exhibit the following interesting features. Firstly, the auxiliary metric $\widetilde{g}_{\mu\nu}$ does not appear by itself in the equations while $\phi$ does explicitly. This means that, by conformal extension of the theory, an extra degree of freedom is obtained which leads to an effectively extra contribution to the stress energy tensor as follows:
\bea
T^{eff}_{\mu\nu}=T_{\mu\nu}+\widetilde{T}_{\mu\nu}=T_{\mu\nu}+(G+T)\partial_{\mu}\phi\partial_{\nu}\phi \label{emt}
\eea
The above equation  can be compared with the energy momentum tensor of the perfect fluid:
\bea
T_{\mu\nu}=(\rho +p)u_{\mu}u_{\nu}-pg_{\mu\nu}\label{perfect}
\eea
where, $\rho$ is energy density, $p$ is pressure, and $u_{\mu}$ is four velocity of the perfect fluid. It is obvious from Eq. (\ref{emt}) that $\widetilde{T}_{\mu\nu}$ is the energy momentum tensor of a perfect fluid with the energy density $G+T$ and zero pressure. $\partial_{\mu}\phi$ represents four velocity of the fluid satisfying  the normalization condition $u_{\mu}u^{\mu}=1$ by Eq. (\ref{constraint}). Eq. (\ref{motion2}) can be interpreted as the conservation of the energy momentum tensor of such a fluid. So, we have a dust with the energy density $G+T$ that does not vanish even in the absence of matter. In this way, this fluid mimics the behavior of dark matter and  it is, therefor, called the mimetic dark matter.
\\An alternative way to formulate mimetic theory is to introduce a scalar field satisfying  constraint (\ref{constraint}) and use the Lagrange multipliers method to obtain equations of motion from the constrained action \cite{lagrange1,lagrange2,lagrange3,lagrange4}:
\bea
S=\int d^{4}x\sqrt{-g}\left[-\frac{1}{2}R(g_{\mu\nu})+\lambda (g^{\mu\nu}\partial_{\mu}\phi\partial_{\mu}\phi -1)\right]\qquad
\eea
Variation with respect to $\lambda$, $g^{\mu\nu}$, and $\phi$ leads to Eqs. (\ref{constraint}), (\ref{motion1}), and (\ref{motion2}), respectively, without the need to introduce any auxiliary metric $\widetilde{g}^{\mu\nu}$.

\section{Teleparallel gravity}
As already mentioned,  Teleparallel gravity is an equivalent formulation of General Relativity that assumes torsion to be responsible for the gravitational interaction \cite{Eins,TEGR}. The fundamental variables in TG are vierbeins or tetrads represented by $\textbf{e}_{A},$ $A=0,1,2,3$. Tetrad fields form an orthogonal basis for the tangent space at each point $x_{\mu}$ of the manifold. Since the tangent space is flat, we have $\textbf{e}_{A}.\textbf{e}_{B}=\eta_{AB}$, where, $\eta_{AB}$ is the Minkowski metric of the flat space time, i.e., $\eta_{AB}=diag(1,-1,-1,-1)$. The vector $\textbf{e}_{A}$ can be described by its components in the coordinate space $\textbf{e}_{A}=e_{A}^{\mu}\partial_{\mu}$ in which the  Latin indices refer to the tangent space and the Greek ones refer to the coordinate space in the manifold. The metric has the following form:
\bea
g_{\mu\nu}=\eta_{AB}e_{\mu}^{A}e_{\nu}^{B}\label{geta}
\eea

The tangent space indices are raised and lowered by $\eta_{AB}$ and the coordinate space ones by $g_{\mu\nu}$. Tetrad fields form an orthogonal basis at each point of the tangent space and the coordinate space; we will, therefore,  have: $e_{\mu}^{A}e_{A}^{\nu}=\delta_{\mu}^{\nu}$ and $e_{\mu}^{B}e_{A}^{\mu}=\delta_{A}^{B}$. Here, we use the curvatureless Weitzenb\"{o}ck connection instead of the torsionless Levi-Civita one:
\bea
\Gamma_{\:\:\:\mu\nu}^{\lambda}\equiv e_{A}^{\lambda}\partial_{\nu}e_{\mu}^{A}.
\eea
As a consequence the covariant derivative of the tetrad field vanishes:
\bea
\nabla_{\nu}e_{\mu}^{A}=\partial_{\nu}e_{\mu}^{A}-\Gamma^{\rho}_{\:\:\:\mu\nu}e^{A}_{\rho}=0
\eea
We define the torsion tensor in terms of the Weitzenb\"{o}ck connection:
\bea
T_{\:\:\:\:\mu\nu}^{\lambda}&=&\Gamma_{\:\:\:\:\nu\mu}^{\lambda}-\Gamma_{\:\:\:\:\mu\nu}^{\lambda}=e_{A}^{\lambda}\left(\partial_{\mu}e_{\nu}^{A}-\partial_{\nu}e_{\mu}^{A}\right)\label{tor}
\eea
Contortion tensor can be defined in terms of torsion tensor as follows:
\bea
K^{\mu\nu}_{\:\:\:\:\:\rho}&=&-\frac{1}{2}\Big(T^{\mu\nu}_{\:\:\:\:\:\rho}-T^{\nu\mu}_{\:\:\:\:\:\rho}-T_{\rho}^{\:\:\mu\nu}\Big)\nonumber\\
\eea
which is the difference between Weitzenb\"{o}ck and Levi-Civita connections:
\bea
K^{\rho}_{\:\:\:\mu\nu}=\Gamma^{\rho}_{\:\:\:\mu\nu}-\bar{\Gamma}^{\rho}_{\:\:\:\mu\nu}\label{k}
\eea
where, $\bar{\Gamma}^{\rho}_{\:\:\:\mu\nu}=\dfrac{1}{2}g^{\rho\sigma}[\partial_{\mu}g_{\nu\sigma}-\partial_{\sigma}g_{\mu\nu}+\partial_{\nu}g_{\sigma\mu}]$ is the Levi-Civita connection. As already mentioned,  the curvature of the Weitzenb\"{o}ck connection vanishes; thus, we have a zero Riemann tensor in terms of the Weitzenb\"{o}ck connection:
\bea
R^{\rho}_{\:\:\:\lambda\mu\nu}=\partial_{\mu}\Gamma^{\rho}_{\:\:\:\lambda\nu}+\Gamma^{\rho}_{\:\:\:\sigma\mu}\Gamma^{\sigma}_{\:\:\:\lambda\nu}-(\mu\leftrightarrow\nu)\equiv 0\label{R}
\eea
Plugging $\Gamma^{\rho}_{\:\:\:\mu\nu}=K^{\rho}_{\:\:\:\mu\nu} +\bar{\Gamma}^{\rho}_{\:\:\:\mu\nu} $ in (\ref{R}), we have:
\bea
R^{\rho}_{\:\:\:\lambda\mu\nu}=\bar{R}^{\rho}_{\:\:\:\lambda\mu\nu}+Q^{\rho}_{\:\:\:\lambda\mu\nu}\equiv 0\label{riman}
\eea
where,
\bea
Q^{\rho}_{\:\:\:\lambda\mu\nu}&=&\overline{\nabla}_{\mu}K^{\rho}_{\:\:\:\lambda\nu}+K^{\rho}_{\:\:\:\sigma\mu}K^{\sigma}_{\:\:\:\lambda\nu}-(\mu\leftrightarrow\nu)\nonumber\\
&=&-\bar{R}^{\rho}_{\:\:\:\lambda\mu\nu}\label{Qriman}
\eea
and $\overline{\nabla}_{\mu}$ is the Levi-Civita covariant derivative:
\bea
\overline{\nabla}_{\rho}V^{\mu}=\partial_{\rho}V^{\mu}+\overline{\Gamma}^{\mu}_{\:\:\:\lambda\rho}V^{\lambda}
\eea
Finally, we define superpotential as in the following:
\bea
S_\rho^{\:\:\:\mu\nu}=\frac{1}{2}\Big(K^{\mu\nu}_{\:\:\:\:\:\rho}+\delta^\mu_\rho \:T^{\alpha\nu}_{\:\:\:\:\:\:\alpha}-\delta^\nu_\rho\:
T^{\alpha\mu}_{\:\:\:\:\:\:\alpha}\Big)\nonumber
\eea
This definition is used to construct the torsion scalar $T$:
\bea
T\equiv S_\rho^{\:\:\:\mu\nu}\:T^\rho_{\:\:\:\mu\nu} \label{T}
\eea
The action of the Teleparallel gravity may then be  defined as follows:
\bea
I=\int d^{4}x e\left(\frac{T}{2}+\mathcal{L}_{m}\right)\label{TGaction}
\eea
where, $e=det(e_{\mu}^{A})=\sqrt{-g}$. The above action is constructed based on the  assumption of invariance under the general coordinate transformation, the local Lorenz transformation, and the parity operation while the Lagrangian density is assumed  to be a second order one in the torsion tensor. The TG action in Eq. (\ref{TGaction}) is equivalent to GR action Eq. (\ref{graction}) up to a total divergence. It only takes a small amount of mathematical calculation to realize
 from Eqs. (\ref{riman}) and (\ref{Qriman}) that
\bea
T=-\bar{R}+2\overline{\nabla}_{\mu}T^{\mu}\label{TR}
\eea
where, $T^{\mu}=T_{\rho}^{\:\:\:\rho\mu}$. A straightforward generalization of TG leads to the following action where $T$ in the action (\ref{TGaction}) is replaced with $f(T)$:
\bea
I=\int d^{4}x e\left(\frac{f(T)}{2}+\mathcal{L}_{m}\right)\label{ftaction}
\eea
Action (\ref{ftaction}) contains all the symmetries of (\ref{TGaction}), except that it lacks the local Lorenz invariance and it is, therefore, not possible to fix some of the field variables by gauge choice \cite{Lorenz}. Varying the action with respect to $e_{\nu}^{A}$ yields the following  equations of motion as:
\bea \label{eom}
J_{A}^{\nu}&\equiv & e^{-1}\partial_{\mu}(ee_A^{\rho}S_{\rho}{}^{\mu\nu})[1+F_{T}(T)]\nonumber\\ &&-e_{A}^{\lambda}T^{\rho}{}_{\mu\lambda}S_{\rho}{}^{\nu\mu}[1+F_{T}(T)]+e_A^{\rho}S_{\rho}{}^{\mu\nu}\partial_{\mu}({T})F_{TT}(T)\nonumber\\ &&-\frac{1}{4}e_{A}^{\nu}[T+F(T)]-\frac{1}{2}e_{A}^{\rho}T^{(m)}_{\:\:\:\rho}{}^{\nu}=0
\eea
where, $f(T)=T+F(T)$ and $T^{(m)}_{\mu\nu}=\dfrac{-2}{e}\dfrac{\delta(e\mathcal{L}_{m})}{\delta g^{\mu\nu}}$ is the energy momentum tensor. The TG equations are obtained by setting $F(T)=0$.\\
If the background metric is assumed to be a flat FRW metric:
\bea
ds^2=dt^2-a^2(t)(dx^i)^2\label{frw}
\eea
then, the vierbeins take the following form:
\bea
e_{\mu}^A=diag(1,a,a,a),\quad e_{A}^{\mu}=diag(1,a^{-1},a^{-1},a^{-1})\quad
\eea
and the torsion scalar becomes
\bea
T=-6H^2\label{TH}
\eea
where, $H$ is the Hubble parameter, $H=\frac{\dot{a}}{a}$. Using the energy momentum tensor of the perfect fluid in Eq. (\ref{perfect}) yields the following  effective Friedman equations:
\bea
H^2&=&\frac{1}{3}\rho-\frac{1}{6}F(T)-2F_TH^2\\
\dot{H}&=&-\frac{\frac{1}{2}(\rho +p)}{1+F_T+2TF_{TT}}\label{ftfrid}
\eea
In what follows, we try to extend the method of mimetic theory to $f(T)$ gravity and to find the equations of motion of dark matter.
\section{Mimetic $f(T)$ teleparallel gravity, variational method}
In this section, the mimetic method of \cite{Mimetic} is applied to TG and $f(T)$ teleparallel gravity. Eq. (\ref{mimetic}) is thus replaced with  an equivalent relation in TG. there is no local Lorenz invariance in the mimetic gravity and this motivates us to study mimetic $f(T)$ gravity. A method to do this is to leave the vierbeins unchanged and to change the Minkowski metric $\eta_{AB}$ as follows:
\bea
\eta_{AB}=\left(\widetilde{\eta}^{CD}\partial_{C}\phi\partial_{D}\phi\right)\widetilde{\eta}_{AB}
=P\widetilde{\eta}_{AB}\label{mimTG}
\eea
To see how this leads to (\ref{mimetic}), we write Eq. (\ref{mimetic}) as follows:
\bea
g_{\mu\nu}&=&\widetilde{g}_{\mu\nu}(\widetilde{g}^{\alpha\beta}\partial_\alpha\phi\partial_\beta\phi) \nonumber\\
&=&\widetilde{\eta}_{AB}e_{\nu}^{A}e_{\mu}^{B}\left(\widetilde{\eta}^{CD}e_{C}^{\alpha}e_{D}^{\beta}\partial_{\alpha}\phi\partial_{\beta}\phi\right)\nonumber\\
&=&e_{\nu}^{A}e_{\mu}^{B}\left[\widetilde{\eta}_{AB}\left(\widetilde{\eta}^{CD}\partial_{C}\phi\partial_{D}\phi\right)\right]\nonumber\\
&=&\eta_{AB}e_{\nu}^{A}e_{\mu}^{B}\nonumber\\
&=&g_{\mu\nu}\nonumber
\eea
where, $\widetilde{g}_{\mu\nu}=\widetilde{\eta}_{AB}e_{\nu}^{A}e_{\mu}^{B}$ is defined in the second line, $e_{\mu}^{A}=\frac{\partial \overline{x}^{A}}{\partial x^{\mu}}$ is used in the third line, and  Eq. (\ref{mimTG}) is used in the forth line. The variation of the metric needs to be determined in  order to vary the action with respect to the new variables $\widetilde{g}_{\mu\nu}$ and $\phi$. At  first sight, the physical metric becomes a function of $\widetilde{\eta}_{AB}$, $\phi$ and $e_{\nu}^{A}$ as follows:
\bea
g_{\mu\nu}=\widetilde{\eta}_{AB}e_{\nu}^{A}e_{\mu}^{B}\left(\widetilde{\eta}^{CD}\partial_{C}\phi\partial_{D}\phi\right)
\eea
It should be noted, however,  that $\eta_{AB}$ is a constant matrix and has zero variation; so:
\bea
\delta g_{\mu\nu}=\delta\eta_{AB}
e_{\nu}^{A}e_{\mu}^{B}+2\eta_{AB}e_{\mu}^{B}\delta e_{\nu}^{A}=2\eta_{AB}e_{\mu}^{B}\delta e_{\nu}^{A}\qquad \label{gg}
\eea
Even so, the term $ \delta\eta_{AB}e_{\nu}^{A}e_{\mu}^{B} $ is not deleted at this juncture  and  its variation is taken into account by using Eq. (\ref{mimTG}):
\bea
\delta g_{\mu\nu}&=&\delta\eta_{AB}
e_{\nu}^{A}e_{\mu}^{B}+2\eta_{AB}e_{\mu}^{B}\delta e_{\nu}^{A}\nonumber\\
&=&P\delta \widetilde{\eta}_{AB}e_{\nu}^{A}e_{\mu}^{B} +\widetilde{\eta}_{AB}e_{\nu}^{A}e_{\mu}^{B} \delta P+2\eta_{AB}e_{\mu}^{B}\delta e_{\nu}^{A}\qquad
\eea
Using the definition of $P$ in (\ref{mimTG}), the variation of $P$ is found to be  $\delta P=-\widetilde{\eta}^{CM}\widetilde{\eta}^{DN}(\delta\widetilde{\eta}_{MN})\partial_{C}\phi\partial_{D}\phi+2\widetilde{\eta}^{CD}(\delta\partial_{C}\phi)\partial_{D}\phi$; thus,
\begin{widetext}
\bea
\delta g_{\mu\nu}&=&P\delta \widetilde{\eta}_{AB}e_{\nu}^{A}e_{\mu}^{B}+\widetilde{\eta}_{AB}e_{\nu}^{A}e_{\mu}^{B}(-\widetilde{\eta}^{CM}\widetilde{\eta}^{DN}(\delta \widetilde{\eta}_{MN})\partial_{C}\phi\partial_{D}\phi+2\widetilde{\eta}^{CD}(\delta\partial_{C}\phi)\partial_{D}\phi)+2\eta_{AB}e_{\mu}^{B}\delta e_{\nu}^{A}\nonumber\\
&=&P\delta \widetilde{\eta}_{MN}e_{\nu}^{A}e_{\mu}^{B}\left(\delta_{A}^{M}\delta_{B}^{N}
-\eta_{AB}\eta^{CM}\eta^{DN}\partial_{C}\phi\partial_{D}\phi\right)
+2\eta_{AB}e_{\nu}^{A}e_{\mu}^{B}\eta^{CD}(\delta\partial_{C}\phi)\partial_{D}\phi+2\eta_{AB}e_{\mu}^{B}\delta e_{\nu}^{A}
\eea
\end{widetext}
In which, the fact that $\widetilde{\eta}_{AB}\widetilde{\eta}^{CD}=\eta_{AB}\eta^{CD}$  was taken into account. To find the variation $ \delta\partial_{C}\phi$, we note that $\delta\partial_{C}\phi=\delta(\partial_{\alpha}\phi e^{\alpha}_{C})= e^{\alpha}_{C}\delta(\partial_{\alpha}\phi)+\partial_{\alpha}\phi \delta(e^{\alpha}_{C})$. Given the fact that $\delta e^{\alpha}_{C}=-g^{\alpha\nu}\eta_{AC}(\delta e_{\nu}^{A})$, the variation of the metric takes the following form:
\begin{widetext}
\bea
\delta g_{\mu\nu}&=&P(\delta \widetilde{\eta}_{MN}e_{\nu}^{A}e_{\mu}^{B}+2\widetilde{\eta}_{MN}e_{\mu}^{B}\delta e_{\nu}^{A})(\delta_{A}^{M}\delta_{B}^{N}-\eta_{AB}\eta^{CM}\eta^{DN}\partial_{C}\phi\partial_{D}\phi)
+2g_{\mu\nu}g^{\alpha\beta}\partial_{\alpha}\delta\phi\partial_{\beta}\phi\nonumber\\
&=&P\delta \widetilde{g}_{\alpha\beta}(\delta_{\mu}^{\alpha}\delta_{\nu}^{\beta}-g_{\mu\nu}g^{\kappa\alpha}g^{\lambda\beta}\partial_{\kappa}\phi\partial_{\lambda}\phi)
+2g_{\mu\nu}g^{\alpha\beta}\partial_{\alpha}\delta\phi\partial_{\beta}\phi \label{vari g}
\eea
\end{widetext}
where, $\eta^{AB}\partial_{A}\phi\partial_{B}\phi=1$, which is equivalent to Relation (\ref{constraint}) as a result of using Eq. (\ref{geta}). Variation of the action (\ref{TGaction}) is given by:
\bea
\delta I&=&2\int d^{4}xeJ_{\nu}^{A}\delta e_{A}^{\nu}=\int d^{4}xeB_{\mu\nu}\delta g^{\mu\nu}\qquad \label{vary I}
\eea
where, the definition of  the tensor $J_{\nu}^{A}$ given in (\ref{eom}) is used and
\bea
&&B_{\mu\nu}=\eta_{AB}e_{\mu}^{B}J_{\nu}^{A}=-\frac{1}{2}\bar{G}_{\mu\nu}-\frac{1}{2}T^{(m)}_{\:\:\:\mu\nu}-S_{\mu\nu}{}^{\alpha}F_{TT}\partial_{\alpha}T\nonumber\\
&&-\frac{1}{2}F_T\left(\bar{R}_{\mu\nu}-g_{\mu\nu}\overline{\nabla}_{\alpha}K^{\alpha}\right)-\frac{1}{4}Fg_{\mu\nu}
\eea
Moreover,  Eq. (\ref{gg})  is used in (\ref{vary I}). By plugging $\delta g^{\mu\nu}$ from (\ref{vari g}) in (\ref{vary I}), we find the following substitute equations for the typical equations of motion of $f(T)$ theory:
\bea
&B_{\mu\nu}-B\partial_{\mu}\phi\partial_{\nu}\phi =0& \label{equ1}\\
&\nabla_{\mu}\left(B\partial^{\mu}\phi\right)=0&\label{equ2}
\eea
in which, $B$ is the trace of tensor $B_{\mu\nu}$ and has the following form:
\bea
B&=&g^{\mu\nu}B_{\mu\nu}=-\frac{1}{2}\bar{G}-\frac{1}{2}T^{(m)}-g^{\mu\nu}S_{\mu\nu}{}^{\alpha}F_{TT}\partial_{\alpha}T\nonumber\\
&&-\frac{1}{2}F_T\left(\bar{R}-4\overline{\nabla}_{\alpha}K^{\alpha}\right)-F
\eea
Where $\bar{G}$ and $T^{(m)}$ are the trace of Einstein and Energy-momentum tensor respectively. Certain points are worthy of mention at this juncture. Eqs. (\ref{equ1}) and (\ref{equ2}) are the mimetic equations for $f(T)$ gravity with no  equivalent  in GR. Setting $F(T)=0$, in (\ref{equ1}), we obtain the mimetic teleparallel gravity, which can be shown to be equivalent to (\ref{motion1}). Hence, the mimetic extensions of TG and GR are similar as expected.\\
 From Eq. (\ref{equ1}), we see that the new degree of freedom leads to an extra contribution to the  energy momentum tensor as follows:
\bea
\widetilde{T}_{\mu\nu}=2B\partial_{\mu}\phi\partial_{\nu}\phi
\eea
Comparison of the above equation with (\ref{perfect}) reveals that $\widetilde{T}^{\mu\nu}$ is equivalent to an effective energy momentum tensor of a perfect fluid with  $\rho=2B$, $p=0$, and four velocity $u^{\mu}=\partial^{\mu}\phi$. It also satisfies the normalization condition $u^{\mu}u_{\mu}=1$ by (\ref{constraint}). Conservation law for $\widetilde{T}_{\mu\nu}$ gives:
\bea
\nabla_{\mu}\widetilde{T}^{\mu}_{\nu}=2\partial_{\nu}\phi\nabla_{\mu}(B\partial^{\mu}\phi)+2B\partial^{\mu}\phi\nabla_{\mu}\partial_{\nu}\phi =0
\eea
which exploits both  Eq. (\ref{equ2}) and the fact that by differentiating (\ref{constraint}) we have $\partial^{\mu}\phi\nabla_{\nu}\partial_{\mu}\phi=
\partial^{\mu}\phi\nabla_{\mu}\partial_{\nu}\phi=0$. In this way, Eq. (\ref{equ2}) becomes the conservation law for $\widetilde{T}^{\mu\nu}$. We see that the scalar field $\phi$ reveals  itself in the equations while the auxiliary metric $\widetilde{g}_{\mu\nu}$ is absent. This means that the  conformal degree of freedom becomes dynamical and acts as a pressureless fluid called `the cold dark matter'.\\
\section{Mimetic $f(T)$ gravity, Lagrange multipliers method}
In the previous section, we discussed the mimetic method for $f(T)$ theory and TG using the variational method with respect to the new variables of the theory $\widetilde{g}_{\mu\nu}$ and $\phi$. Its requisite is the definition of the physical metric $g_{\mu\nu}$ in terms of the auxiliary metric $\widetilde{g}_{\mu\nu}$ and $\phi$ in Eq. (\ref{mimetic}) or equivalently (\ref{mimTG}). In this section we present a simpler method that is based on  constraint (\ref{constraint}). If we implement this constraint or equivalently either Eq. (\ref{mimetic}) or (\ref{mimTG}) in the action using the undetermined Lagrange multipliers method, we  obtain similar results. We define the action as:
\bea
&&I=\int d^{4}x\:e \label{consaction}\\
&&\left[\frac{1}{2}\left(T+F(T)\right)+\lambda^{\mu\nu}\left(\widetilde{g}_{\mu\nu}\widetilde{g}^{\alpha\beta}\partial_\alpha\phi\partial_\beta\phi -g_{\mu\nu}\right)+\mathcal{L}_{m}\right]\nonumber
\eea
Variation of Eq. (\ref{consaction}) with respect to $\lambda^{\mu\nu}$ leads to the constraint (\ref{mimetic}). Moreover, its variation  with respect to $\widetilde{g}_{\mu\nu}$ yields the following equation:
\bea
\lambda^{\mu\nu}\widetilde{g}^{\alpha\beta}(\partial_{\alpha}\phi)(\partial_{\beta}\phi)-\lambda^{\rho\sigma}\widetilde{g}_{\rho\sigma}\widetilde{g}^{\mu\alpha}(\partial_{\alpha}\phi)\widetilde{g}^{\nu\beta}(\partial_{\beta}\phi)=0\qquad\label{gtil}
\eea
(\ref{gtil}) yields (\ref{lambda}) below if (\ref{mimetic}) is used:
\bea
\lambda_{\mu\nu}=\lambda(\partial_{\mu}\phi)(\partial_{\nu}\phi)\label{lambda}
\eea
where, $\lambda$ is the trace of  $ \lambda_{\mu\nu} $. Variation with respect to $g_{\mu\nu}$ leads to:
\bea
B^{\mu\nu}-\lambda^{\mu\nu}=0\label{g}
\eea
Taking trace of the above equation gives:
\bea
B-\lambda =0\label{lambda2}
\eea
Replacing Eqs. (\ref{g}) and (\ref{lambda2}) in (\ref{lambda}) yields Eq. (\ref{equ1}). Finally, variation with respect to $\phi$ leads to:
\bea
\nabla_{\mu}(\lambda\partial^{\mu}\phi)=0\label{phi}
\eea
This is Eq. (\ref{equ2}) in which $\lambda$ is replaced from (\ref{lambda2}).\\
Similar to what is claimed in \cite{lagrange2}, since $\lambda_{\mu\nu}$  is fully determined by its trace, we can implement the following constraint in the action for GR as well and write (\ref{consaction2}) below instead of (\ref{consaction}):
\bea
I=\int d^{4}xe\left[\frac{1}{2}\left(T+F(T)\right)+\lambda\left(g^{\mu\nu}\partial_{\mu}\phi\partial_{\nu}\phi  -1\right)+\mathcal{L}_{m}\right]\nonumber\\\label{consaction2}
\eea
Eqs. (\ref{constraint}), (\ref{equ1}), and (\ref{equ1}) are obtained by varying the action (\ref{consaction2}) with respect to $\lambda$, $g^{\mu\nu}$ and $\phi$.
\section{Mimetic $f(T)$ gravity and dynamical systems}
\begin{table*}
\begin{center}
\vspace{5mm}
\begin{tabular}{|c|c|c|c|c|c|c|c|c|c|}
\hline

{\bf fixed point} & $x$ & $\Omega_{m}$ & $\Omega_{r}$&$\Omega_{dm}$& $\Omega_{de}$&$q$&$\lambda_{1}$ & $\lambda_{2}$&$\lambda_{3}$   \\\hline\hline

$P_{1}$& $x_0$& $1+x_0$ & $0$ & $0$&$-x_0$ &$\frac{1}{2}$&$0$&$ -1$&$\frac{3}{2}(y'(x_0)+2)$    \\\hline

$P_{2}$&$x_0$ & $0$ & $1+x_0$ & $0$& $-x_0$&$1$&$1$&$1$&$2\left(y'(x_0)+2\right)$  \\\hline

$P_{3}$&$x_0$ & $0$ & $0$ & $1+x_0$& $-x_0$&$\frac{1}{2}$&$0$&$-1$&$\frac{3}{2}\left(y'(x_0)+2\right)$  \\\hline

$P_{4}$&$x_1$ & $0$ & $0$ & $0$& $1$&$-1$& $-3$&$-4$&$-3$   \\\hline

$P_{5}$&$x_2$ & $0$ & $0$ & $1-x_2-y(x_2)$& $x_2+y(x_2)$&$-1$&$-3$&$-4$&$\lambda (x_2)$   \\\hline

\end{tabular}
\caption{ The critical points and related physical parameters of the system (\ref{ode}).The solution of $y(x)+2x=0$ is called $x_0$, the solution of $y(x)+x=1$ is called $x_1$, and the value in which $1-\frac{1}{2}y+y'\left(x+\frac{1}{2}y\right)$ diverges is called $x_2$.}

\label{tab}
\end{center}
\end{table*}
The method of dynamical systems is a tool for investigating the whole dynamics of a theory \cite{rev fix}. In this method qualitative behavior of the system is studied near the extremums of the theory called fixed points or critical points. Based on different initial values there could be different solutions. Inconsistent solutions should be ruled out regarding early and late time behaviors of the universe, and matter/radiation solutions. In this section we investigate the method of autonomous dynamical systems to examine the mimetic $f(T)$ theory and see if it is a viable cosmological theory or not. We consider $f(T)$ mimetic gravity in flat homogeneous and isotropic FRW universe. The content of the universe is chosen to be dust (with energy density $\rho_m$, and pressure $p_m=0$) and radiation (with density $\rho_r$ and pressure $p_r=\frac{1}{3}\rho_r$). The action reads:
\bea
I&=&\int d^{4}xe[\frac{1}{2}\left(T+F(T)\right)+\lambda\left(g^{\mu\nu}\partial_{\mu}\phi\partial_{\nu}\phi -1\right) +\mathcal{L}_{m} \nonumber\\
&&\qquad\quad+\mathcal{L}_{r}]\label{consaction3}
\eea
Then, variation with respect to $e_{\nu}^{A}$ (or equivalently $g^{\mu\nu}$) leads to following equations of motion:
\bea
&&e^{-1}\partial_{\mu}(ee_A^{\rho}S_{\rho}{}^{\mu\nu})[1+F_{T}(T)]-e_{A}^{\lambda}T^{\rho}{}_{\mu\lambda}S_{\rho}{}^{\nu\mu}(1+F_{T}(T))\nonumber\\
&&-\frac{1}{2}e_{A}^{\rho}T^{(m)}_{\:\:\:\rho}{}^{\nu}-\frac{1}{2}e_{A}^{\rho}T^{(r)}_{\:\:\:\rho}{}^{\nu}+e_A^{\rho}S_{\rho}{}^{\mu\nu}\partial_{\mu}({T})F_{TT}({T})\nonumber\\
&&-\frac{1}{4}e_{A}^{\nu}(T+F(T))-\lambda (\phi)\partial_{\mu}\phi\partial^{\nu}\phi e^{\mu}_{A}=0
\eea
and variation with respect to $\phi$ gives:
\bea
2\nabla_{\mu}(\lambda (\phi)\partial^{\mu}\phi)=2\frac{1}{a^3}\partial_{\mu}(a^3\lambda \partial^{\mu}\phi)=0\label{contin}
\eea
Based on the homogeneity and isotropy of the universe $\lambda(\phi)$ which is equal to the density of the mimetic mater should depends only on time. So we assume that the scalar field $\phi$ depends only on time. Regarding Eq. (\ref{frw}), Eq. (\ref{constraint}) then changes to:
\bea
\left(\frac{d\phi}{dt}\right)^2=1
\eea
Thus, we can identify the auxiliary field $\phi$ by $t$. In the flat FRW universe with metric (\ref{frw}), the effective Friedman equations would then become:
\bea
3H^2(1+2F_T)+\frac{1}{2}F(T)&=&\rho_m +\rho_r +2\lambda \label{fri1}\\
-2\dot{H}(1+F_T+2TF_{TT})&=&\rho_m +\frac{4}{3}\rho_r +2\lambda \label{fri2}\qquad
\eea
The continuity equations and Eq. (\ref{contin}) read:
\bea
&\dot{\rho}_m+3H\rho_m =0&\label{contin1}\\
&\dot{\rho}_r+4H\rho_r =0&\label{contin2}\\
&2\dot{\lambda}+6H\lambda =0&\label{contin3}
\eea
Eqs. \eqref{fri1} to \eqref{contin3} define the whole dynamics of the theory. To simplify the calculations we introduce the following dimensionless variables:
\bea
\Omega_{m}&=&\frac{\rho_m}{3H^2}\label{omegam}\\
\Omega_{r}&=&\frac{\rho_r}{3H^2}\label{omegar}\\
\Omega_{dm}&=&\frac{2\lambda}{3H^2}\label{omegadm}\\
x&=&\frac{F(T)}{T}\label{x}\\
y&=&-2F_T(T)\label{y}\\
z&=&2TF_{TT}(T)\label{z}
\eea
The term $2\lambda$ in Eq. \eqref{fri1} is equal to the effective density of the mimetic dark matter. So we defined the density parameter of the mimetic dark matter as $ \Omega_{dm}=\frac{2\lambda}{3H^2} $ in Eq. \eqref{omegadm}. In the same way using Eq. \eqref{fri1} the effective density of dark energy is $\rho_{de}=-\frac{F(T)}{2}-6H^2F_T(T)$, therefore the density parameter of the effective dark energy,  using Eq. \eqref{TH}, will become $\Omega_{de}=x+y$. Based on the variables \eqref{omegam} to \eqref{z}, Eqs. \eqref{fri1} and \eqref{fri2} will become:
\bea
&&\Omega_m+\Omega_r +\Omega_{dm}+x+y=1\label{ffr1}\\
&&\frac{\dot{H}}{H^2}\left(1-\frac{1}{2}y+z\right)=-\frac{3}{2}\Omega_{m}-\frac{3}{2}\Omega_{dm}-2\Omega_r \qquad \label{ffr2}
\eea
In order to construct an autonomous dynamical system we should select an independent set of the variables from \eqref{omegam} to \eqref{z}. Clearly $x$, is a function of $T$, so we can write $T=T(x)$. In this way $y(T)$ and $z(T)$ are functions of $x$ and cannot interpreted as independent variables of $x$. An explicit relation could be found between $y$ and $z$ by differentiating $y$ with respect to $x$:
\bea
y'(x)=  \frac{dy}{dx}=  \frac{dy/dT}{dx/dT}=  \frac{-2F_{TT}(T)}{\frac{F_T(T)}{T}-\frac{F(T)}{T^2}}=  \frac{z(x)}{x+\frac{1}{2}y(x)}\label{y'}\qquad
\eea
It should be noted that, $y'$ is differentiation of $y$ with respect to $x$, not $T$. Equation \eqref{ffr1} is another constraint between the remained quantities \eqref{omegam} to \eqref{x}. It means that there are three independent variables. We choose $\Omega_m$, $\Omega_r$ and $x$ as independent variables. The autonomous dynamical system using Eqs. \eqref{contin1}, \eqref{contin2}, \eqref{contin3}, \eqref{ffr1}, \eqref{ffr2} and \eqref{y'} will take the following form:
\bea
\frac{d\Omega_m}{dN}&=&\Omega_m\left(\frac{3(1-x-y)+\Omega_r}{1-\frac{1}{2}y+y'\left(x+\frac{1}{2}y\right)}-3\right)\nonumber\\
\frac{d\Omega_r}{dN}&=&\Omega_r\left(\frac{3(1-x-y)+\Omega_r}{1-\frac{1}{2}y+y'\left(x+\frac{1}{2}y\right)}-4\right)\nonumber\\
\frac{dx}{dN}&=&\left(x+\frac{1}{2}y\right)\frac{3(1-x-y)+\Omega_r}{1-\frac{1}{2}y+y'\left(x+\frac{1}{2}y\right)}\label{ode}
\eea
where, $N=\ln a$. The fixed points of the system could be achieved by setting $\frac{dx}{dN}=  \frac{d\Omega_r}{dN}= \frac{d\Omega_m}{dN} =0$. Assuming $1-\frac{1}{2}y(x)+y'(x)\left(x+\frac{1}{2}y(x)\right)=  1+F_T(T)+2TF_{TT}(T)\neq 0$ five categories of fixed points can be found which are summarized in Table \ref{tab}. Related physical parameters are presented in the Table as well. Stability state of the fixed points can be studies with calculating the eigenvalues of the Jacobian matrix near the critical points which are designated by $\lambda_i$ in Table \ref{tab}. If all the eigenvalues are positive, the point is unstable. If all of them are negative the point is stable, and otherwise it is a saddle fixed point. Nonhyperbolic solutions occur when some or all of the eigenvalues have zero real part near the fixed point. Deceleration parameter $q$, presented in the Table is defined as below:
\bea
q=  -\frac{a\ddot{a}}{\dot{a}^2}=  -1-\frac{\dot{H}}{H^2}=  -1+\frac{\frac{3}{2}(1-x-y)+\frac{1}{2}\Omega_r}{1-\frac{1}{2}y+y'\left(x+\frac{1}{2}y\right)}\nonumber\\
\eea
Accelerating universe corresponds to minus values of the deceleration parameter.
\begin{figure*}
\begin{center}
\includegraphics[scale= 0.8]{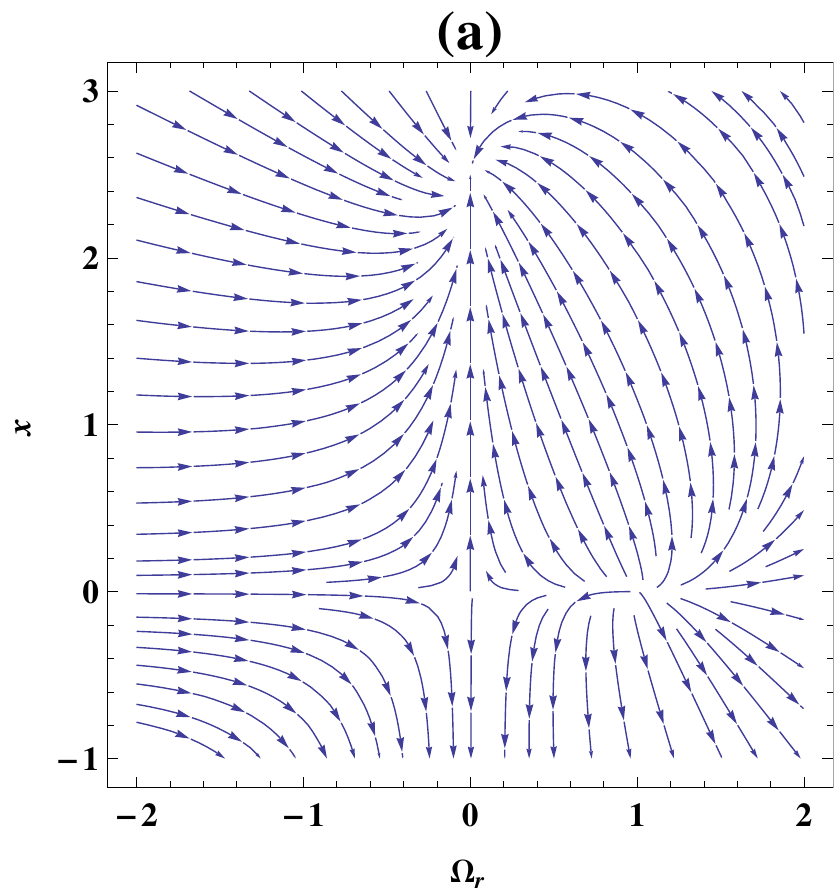}\hspace{10mm}\includegraphics[scale=0.8]{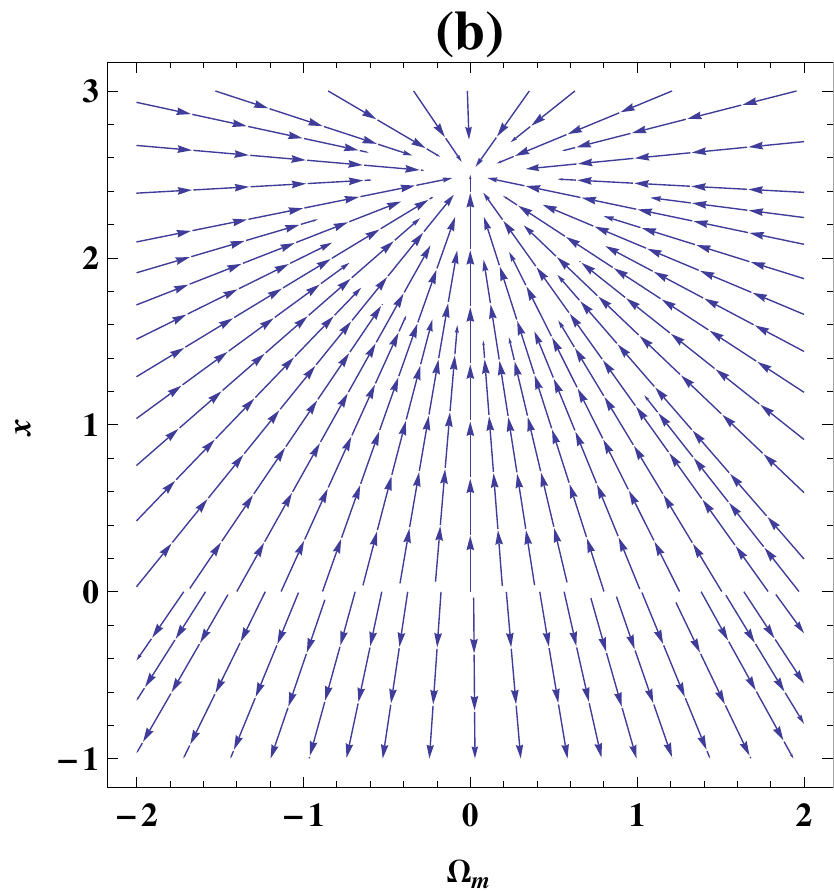}\\\includegraphics[scale=0.8]{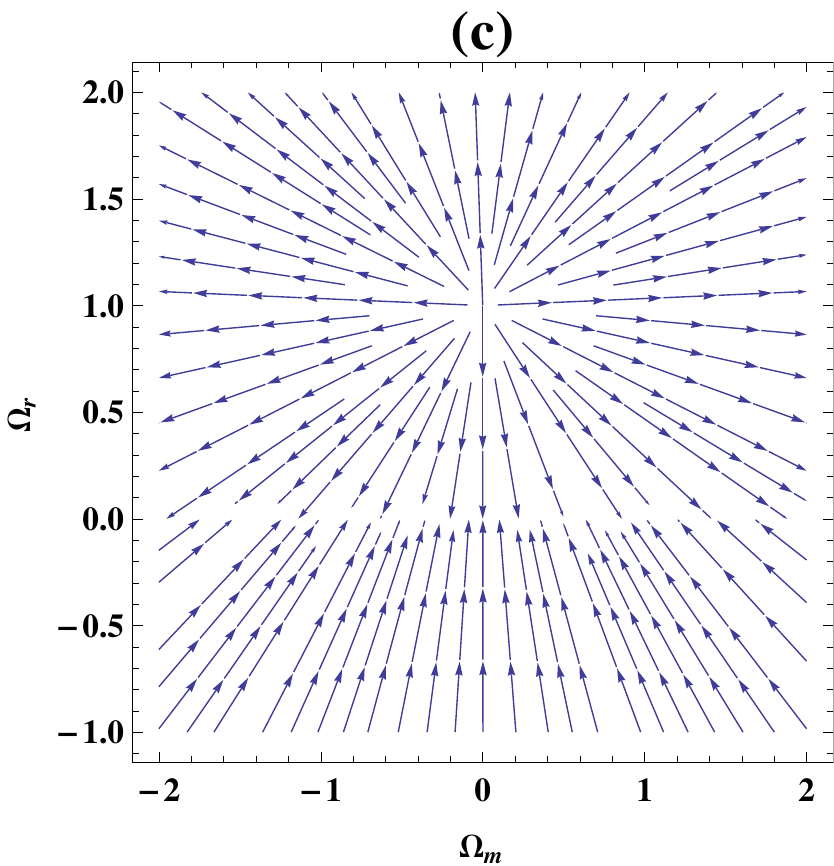}
\caption{ Phase trajectories for the system (\ref{ode}) for $F(T)=\alpha(-T)^b$ where $b=0.3$, an arbitrary $\Omega_m$ in plot \textbf{(a)}, $\Omega_r=0$ in plot \textbf{(b)} and $x=0$ in plot \textbf{(c)} has been chosen. The critical point $P_2$ is located at $(1,0)$ in plot \textbf{(a)} and $(0,1)$ in plot \textbf{(c)}. The critical point $P_4$ is located at $(0,\frac{1}{1-2b})$ in both plots \textbf{(a)} and \textbf{(b)}. The critical line $P_{13}$ is located at $(0,0)$ in plot \textbf{(a)}, at $x=0$ in plot \textbf{(b)} and at $\Omega_r=0$ in plot \textbf{(c)}. $P_4$ is clearly a stable fixed point because the trajectories are evolving through it. $P_{13}$ line is a saddle one because some trajectories evolve through and some of them come out of the line. Finally $P_2$ is an unstable fixed point because all of the trajectories are coming out of it. The trajectories start their evolution from the radiation point $P_2$, the only unstable fixed point in the theory. Some of the trajectories go straight forward to $P_4$, the stable point and some of them evolve first to matter-dark matter critical line $P_{13}$, and after that to $P_4$ (dark energy era), the only attractor in the theory.}
\label{fig1}
\end{center}
\end{figure*}
\\Our universe is in the dark energy dominated accelerated expanding phase. In the past it was smaller and denser than it is today and it was a period of time that matter had been dominated the universe. Since the rate of expansion of radiation is faster than matter in an accelerating universe (based on the solution of equation \eqref{contin2}) before matter, radiation had been dominated the universe. Before radiation and after big bang it is believed that there had been an accelerated expanding era, called inflation. In this way there are four main cosmological periods in standard cosmological model that are inflation, radiation, matter and dark energy dominated eras respectively. Every viable cosmological model should describe some of them, specially the nowadays accelerating phase of the universe. For this goal inflation should be an unstable fixed point and radiation and matter eras should be saddle ones. At last dark energy dominated era should be a stable critical point.
\\The first three categories of fixed points occur with the solutions of $y(x)+2x=0$. The solutions of this equation is called $x_0$. Depending on the functionality of $y(x)$ or $F(T)$, it is possible to have one, two, more or zero fixed points for each category. The first category ($P_1$ in the table) are those for which $\Omega_m=1+x_0$, $\Omega_r=\Omega_{dm}=0$ and $\Omega_{de}=-x_0$. It changes to a matter-dominated fixed point for small values of $x_0$ because $\Omega_m=1$. Also the point is a nonhyperbolic fixed point (there is one zero eigenvalue) but it is possible to understand that it is a saddle fixed point for $y'(x_0)>-2$. $P_1$ changes to a dark energy-dominated fixed point for $x_0=-1$. Since there is no accelerating solution at this case ($q>0$), it fails to describe the late time accelerating phase of the universe.
\\The second category of fixed points ($P_2$ in the table) occur with $\Omega_r=1+x_0$, $\Omega_m=\Omega_{dm}=0$ and $\Omega_{de}=-x_0$. For small values of $x_0$, $P_2$ is a radiation-dominated fixed point. It is an unstable fixed point for $y'(x_0)>-2$, and a saddle one otherwise. Again $P_2$ changes to a dark energy-dominated fixed point for $x_0=-1$. Since there is no accelerating and stable solution, it fails to describe the late time accelerating phase of the universe.
\\The third category of fixed points ($P_3$), are those in which mimetic dark matter dominates if $x_0=0$. It is a saddle fixed point for $y'(x_0)>-2$. The point changes to a dark energy dominated fixed point for $x_0=-1$ but for lack of accelerating solution it cannot describe the late time accelerating phase of the universe.
\\The forth category ($P_4$), are the solutions of $y(x)+x=1$. The roots of this equation are called $x_1$. It is a dark energy dominated, stable, accelerating fixed point and it can describe the late time accelerating phase of the universe correctly.
\\The fifth category ($P_5$), are those for which $1-\frac{1}{2}y+y'\left(x+\frac{1}{2}y\right)$ diverges and $\Omega_{de}\neq 1$. These points are called $x_2$. Existence of nonzero density parameter of dark matter in an accelerating phase ($q<0$) demonstrates the existence of inflation dominated era. It is a saddle fixed point if $\lambda (x)>0$, and a stable one otherwise where, $\lambda (x)$ has the following form:
\bea
\lambda (x)&=  &\frac{3}{2}\frac{(y'+2)(1-x-y)-(y'+1)(y+2x)}{1-\frac{1}{2}y+y'\left(x+\frac{1}{2}y\right)}\\
&-&\frac{3}{2}\frac{(1-x-y)(y+2x)\left(\frac{1}{2}y'(y'+1)+y''\left(x+\frac{1}{2}y\right)\right)}{\left( 1-\frac{1}{2}y+y'\left(x+\frac{1}{2}y\right)\right) ^2}.\nonumber
\eea
In the standard cosmological model discussed above inflation should be an unstable fixed point. Since it is not possible for $P_5$ to be an unstable fixed point it cannot assumed as a true inflation point.
\\In our theory we have an inflation, radiation, ordinary matter and mimetic dark matter-dominated fixed points if $x_0$ tends to zero. As already discussed, $P_5$ cannot be accepted as a true inflation point and it should be ruled out. There is no situation for which both matter and radiation fixed points are saddle ones. But if we choose $f(T)$ in a way that $y'(x_0)>-2$, there will be an unstable radiation era followed by a saddle matter and dark matter fixed points and a stable dark energy one in the theory. This case has an advantage that the transition radiation$\longrightarrow$matter$\longrightarrow$dark energy in the standard cosmological model will becomes possible. Based on the energy conditions, the true domain of $\Omega_r$ and $\Omega_m$ are $0\leq\Omega_r\leq 1$ and $0\leq\Omega_m\leq 1$. Since the density parameters of the matter and radiation in Table \ref{tab}, are equal to $1+x_0$, the preferred domain is, $0\leq 1+x_0\leq 1$ or $-1\leq x_0\leq 0$. For the lower bound, $x_0=-1$, the three first category of fixed points are dark energy dominated fixed points. Since the value of deceleration parameter is positive for all of them, they are in the decelerating phase and relates to no physical situation. Then the preferred domain is $-1< x_0\leq 0$. Thus every viable cosmological model of mimetic $f(T)$ gravity should satisfy some conditions: first, the model has at least one fixed point for each of the categories of Table \ref{tab}. In other words, both of the equations, $y(x)+x=1$ and $y(x)+2x=0$ have at least one real solution. Second, for satisfaction of energy conditions, $-1< x_0\leq 0$, and third, for stability states, $y'(x_0)>-2$. We conclude that, $f(T)$ mimetic theory, with some conditions could be a viable cosmological model.
\\In a theory, if the domain of the dynamical variables are noncompact (i.e. it is possible for the dynamical variables to tend to infinity), the study of the phase space completes with studying the infinite fixed points. In our theory based on energy conditions the dynamical variables $\Omega_m$ and $\Omega_r$ are restricted to $0\leq\Omega_r\leq1$ and $0\leq\Omega_m\leq1$, so they are always finite. $x$ is the only dynamical variable which may tend to infinity. If in the valid domain of $T$, which is determined with the energy conditions, $x(T)$ tends to infinity, it is necessary to study the infinite critical points using the Poincar\'{e} central projection method \cite{poincare}.\\
\\To see an explicit example we use the power-law model of the following form \cite{powerlaw}:
\bea
F(T)=\alpha(-T)^b, \label{powerlaw}
\eea
where $\alpha$ and $b$ are two constants. As we obtained in Ref. \cite{Oboudiat} the preferred values of the independent model parameter to satisfy the conditions are $b <1$ and $b\neq \frac{1}{2}$. The fixed points of the power-law model is plotted in Figure \ref{fig1} for $b=0.3$. Since both equations $y(x)+2x=0$ and $y(x)+x=1$ have real solutions for $b <1$ and $b\neq \frac{1}{2}$, the radiation, matter and dark energy fixed points exist. But there is no quantity for which $1-\frac{1}{2}y+y'\left(x+\frac{1}{2}y\right)$ diverges, so there is no inflation point, $P_5$, in the theory. For the power-law model, when $\Omega_r=x=0$ the first equation of \eqref{ode} vanishes for all values of $\Omega_m$. This leads to a fixed line $P_{13}$ instead of two fixed points $ P_1$ and $P_3$ that is observable in plots \textbf{(b)} and \textbf{(c)} in Figure \ref{fig1}.

\section{Concluding remarks}
In this work, we generalized the idea of mimetic theory to the $f(T)$ teleparallel gravity. The idea of mimetic theory is to extend the metric in terms of an auxiliary conformal metric and a scalar field appearing through its first derivative. It is interesting that the conformal extension of the physical metric acts completely on the Minkowski metric of tangent space and leaves the vierbeins unchanged in the teleparallel gravity. This results in the splitting of the  $f(T)$ gravity equations of motion  into two groups of equations obtained through variation with respect to the auxiliary metric and the scalar field. Thus, the scalar field or the conformal degree of freedom reveals by itself in the equations to become dynamical. This extra degree of freedom acts as a pressureless fluid that can mimic the behavior of cold dark matter. The equations of motion can be obtained in a  simpler  way by implementing Eq. (\ref{mimetic}) as a constraint in the action. In order to examine the viability of the theory, the method of dynamical systems is used. It is found that it is possible to have five categories of fixed points representing inflation, radiation, ordinary matter, mimetic dark matter and dark energy dominated eras. Finally one special case that is power-law model is studied with its phase trajectories.

\end{document}